\documentclass{PoS}
\usepackage{lineno}

\title{Search for dark matter with metastable mediators with the IceCube observatory}

\ShortTitle{A}

\author{
The IceCube Collaboration\footnote{For collaboration list, see PoS(ICRC2019)1177}\\
{\itshape \href{http://icecube.wisc.edu/collaboration/authors/icrc19_icecube}{http://icecube.wisc.edu/collaboration/authors/icrc19\_icecube}}\\
E-mail: \email{ctoennis@icecube.wisc.edu}
}

\abstract{

The IceCube neutrino observatory is a 3D array of photodetectors installed in the Antarctic ice. It consists of 5,160 photomultiplier-tubes spread among 86 vertical strings making a total detector volume of more than a cubic kilometer. It detects neutrinos via Cherenkov light of charged relativistic particles from neutrino interactions with the detector volume. IceCube is, due to its size and photosensor spacing, particularly sensitive to high-energy neutrinos. In this analysis we search for dark matter that annihilates into a metastable mediator that subsequently decays into Standard Model particles. These models yield an enhanced high-energy neutrino flux from dark matter annihilation inside the Sun compared to models without a mediator. Neutrino signals that are produced directly inside the Sun are strongly attenuated at higher energies due to interactions with the solar plasma. In the models considered here, the mediator can escape the Sun before producing any neutrinos, thereby avoiding attenuation. IceCube is ideal to search for this enhanced high-energy neutrino signal. We present the sensitivities of an analysis of six years of IceCube data looking for dark matter in the Sun considering mediator lifetimes between 1 ms to 10 s and dark matter masses between 200 GeV and 10 TeV. We show that IceCube is sensitive to spin--dependent cross--sections of $3.45 \times 10^{-34}~\rm cm^2$ for dark matter masses of 1 TeV.\\

\vspace{4mm}
{\bfseries Corresponding authors:}
\speaker{Christoph T\"onnis}$^{1}$\\
{$^{1}$\itshape Department of Physics, Sungkyunkwan University, Suwon 16419, Korea}\\

}

\FullConference{36th International Cosmic Ray Conference -ICRC2019-\\
		July 24th - August 1st, 2019\\
		Madison, WI, U.S.A.}

\begin{document}

\section{IceCube}\label{sec:IC}

The IceCube Neutrino Observatory~\cite{detector} is to date the largest neutrino telescope. Located at the geographic South Pole, the observatory consists of one cubic kilometer--scale Cherenkov radiation detector built in ice. IceCube has a cubic kilometre of instrumented volume at depths between 1.450~km and 2.450~km, as well as a square kilometer large cosmic--ray air--shower detector at the surface of the ice. The primary scientific goals of the detector are to measure high-energy astrophysical neutrinos and to identify their sources. The IceCube collaboration reported the first observation of an astrophysical neutrino flux~\cite{HESE2013} in 2013, and in 2018, the detector found evidence of a flaring blazar as a source of the astrophysical neutrinos~\cite{TXS}. Other scientific goals of the detector include measurement of neutrino oscillation parameters, cosmic--ray flux, and search for dark matter and other exotic particles. \\

\section{Secluded dark matter}\label{sec:Model}

Secluded dark matter is a particular model of dark matter where the dark matter particle decays or annihilates into a metastable mediator particle before yielding any matter particles~\cite{SDM_basic}. There are a variety of different scenarios for dark matter with such a mediator. Various supersymmetric models of dark matter exist where some superpartner particle is a mediator~\cite{SDM_SUSY}. Alternatively there are models where the mediator is a dark photon~\cite{SDM_Photon} or a dark higgs particle~\cite{SDM_Higgs}. 

\begin{figure}
	\centering
	\includegraphics[width=0.7\textwidth]{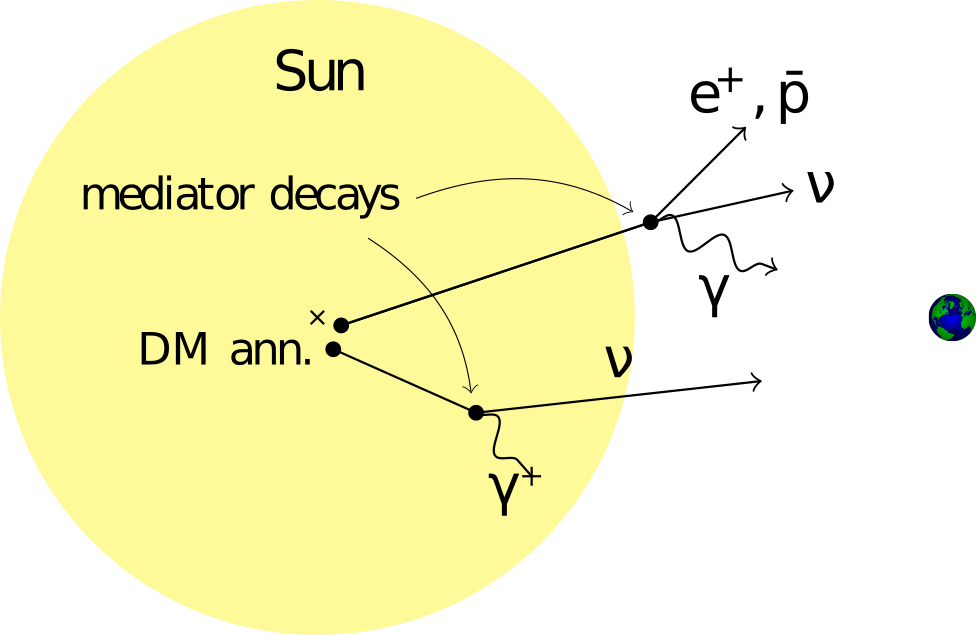}
	\caption{A diagram of secluded dark matter annihilations in the Sun. Two mediators paths are shown: one with a decay length larger than the Sun radius and another mediator decaying inside the Sun.}\label{fig:diagram}
\end{figure}

These mediators are assumed to have only negligible interactions with standard model particles and would be able to propagate through the solar plasma without any significant amount of scattering or absorption. If the mediator is much lighter than the dark matter particle it will move at a highly relativistic speed and escape the Sun after  $\frac{\rm R_{Sun}}{ \rm c} \sim$ 2.3~s and produce standard model particles outside of the plasma. Standard model signals from these mediators are, therefore, not affected by absorption in the Sun. For neutrino signals this would particularly enhance the high energy spectrum of neutrinos, above 1~TeV, since as the Sun becomes opaque to neutrinos of these energies. A diagram of this process is shown in figure \ref{fig:diagram}.

\begin{figure}
	\centering
	\includegraphics[width=0.7\textwidth]{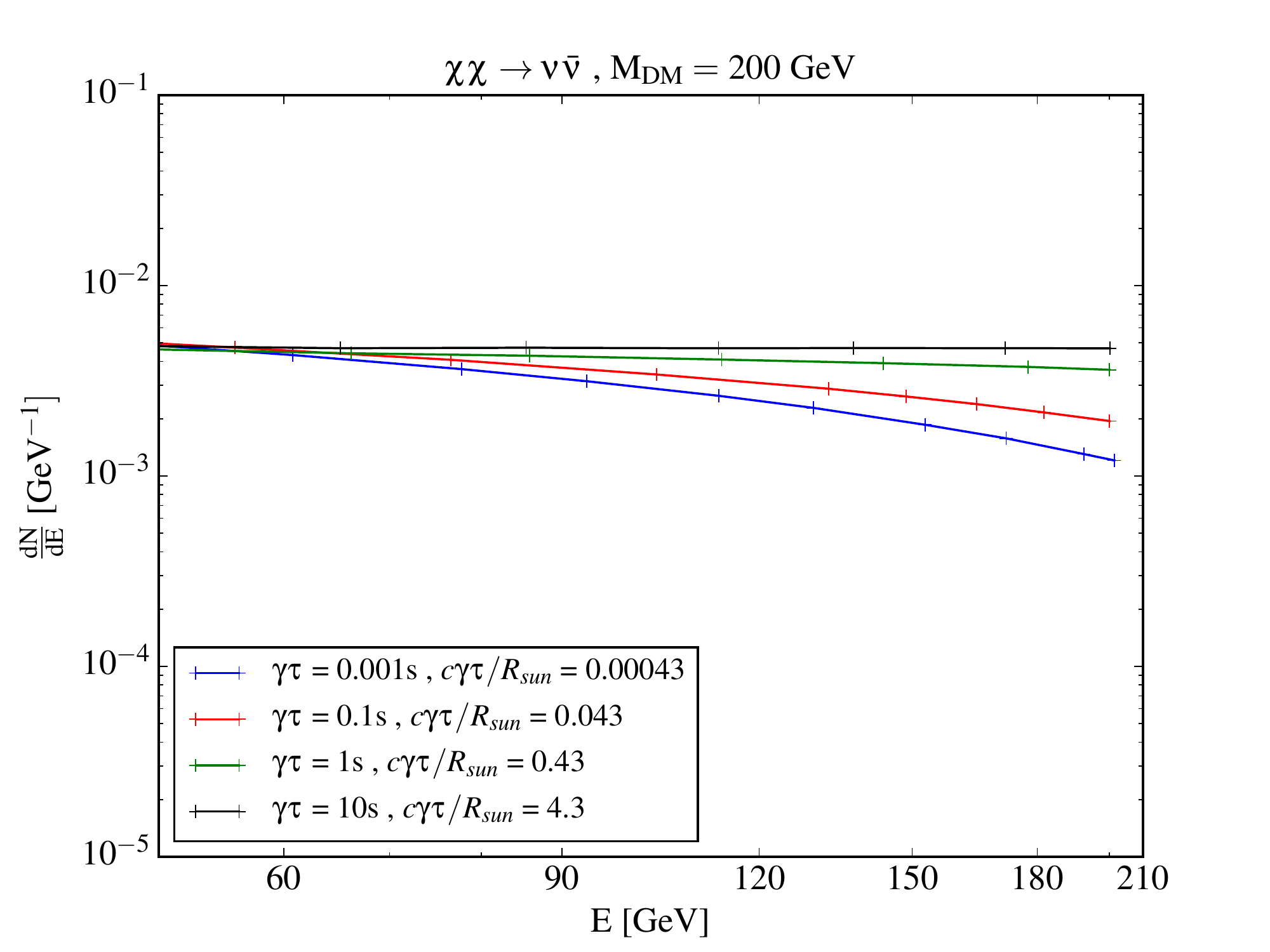}
	\caption{Muon neutrino spectra for a 200 GeV dark matter particle assuming all mediators always decay into neutrinos.}\label{fig:200G}
\end{figure}

\begin{figure}
	\centering
	\includegraphics[width=0.7\textwidth]{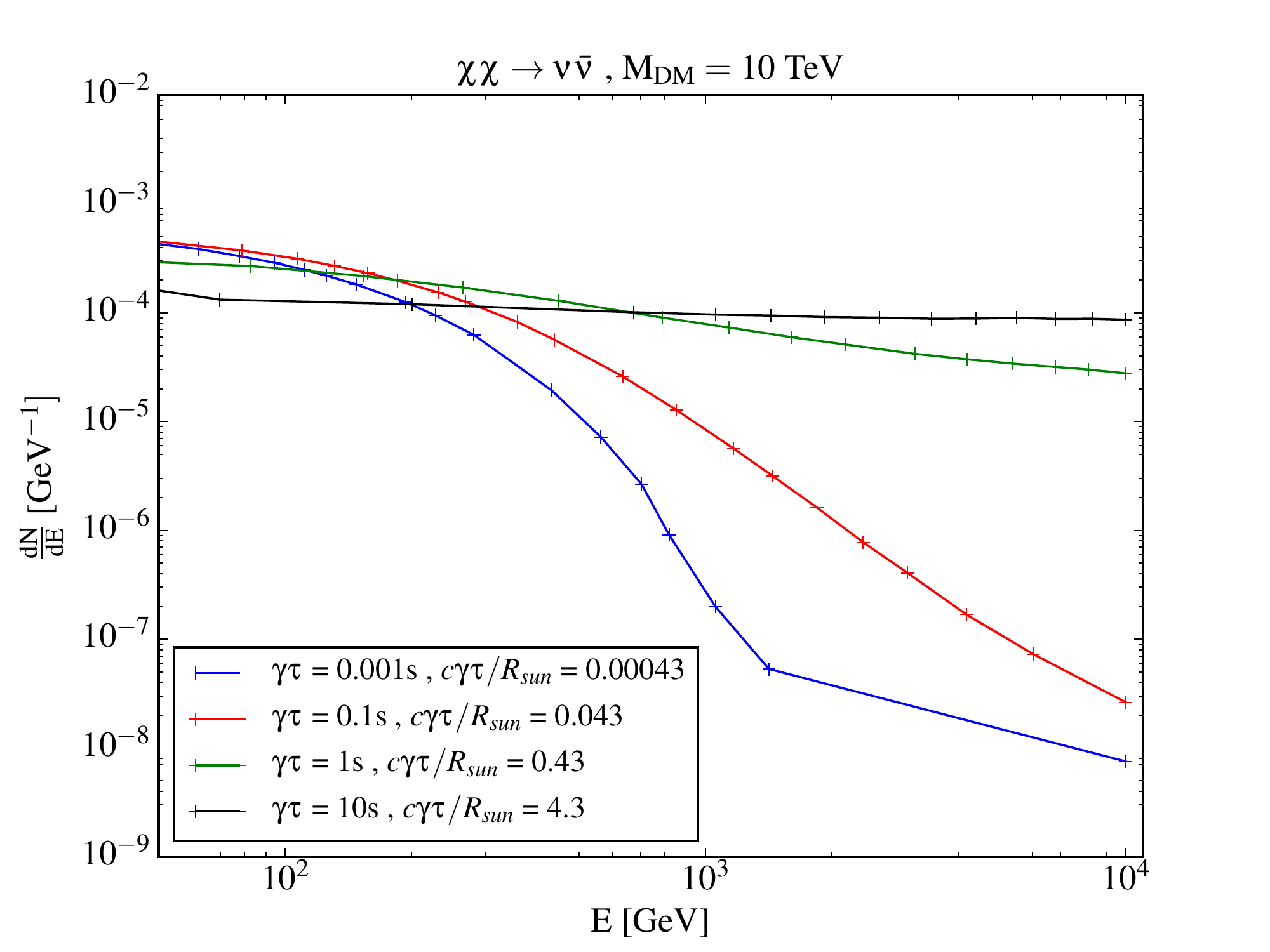}
	\caption{Muon neutrino spectra for a 10 TeV dark matter particle assuming all mediators always decay into neutrinos.}\label{fig:10T}
\end{figure}

The effect of this lack of absorption can be seen in figures \ref{fig:200G} and \ref{fig:10T}. The neutrino spectra shown there display a reduced number of high energy neutrinos for shorter mediator lifetimes, particularly at higher dark matter particle masses one can see an effective cutoff of the signal at energies above 1~TeV due to neutrino absorption. At lower energies more neutrinos can be found for shorter lifetimes, as neutrinos that merely scatter in the solar plasma can lose energy while still contributing to the total neutrino flux.

This amplified high energy neutrino flux makes neutrino telescopes such as IceCube especially sensitive to secluded dark matter. A previous external study using IceCube public sensitivities~\cite{SDM_basic}, was published using sensitivities for 317 live--days of data showing the potential of IceCube to search for secluded dark matter. In this analysis we use 1058 live--days of data and simulation estimates of IceCube. 

\section{The analysis method}\label{sec:Method}

For this analysis a sample of muon neutrinos recorded from 2011 to 2016 with around 1058 days of livetime. This sample only contains muons that travel upwards in the detector to exclude a substantial background of atmospheric muons. This sample has been previously used in other analyses~\cite{Reimann}.\\
In this search a wide range of model parameters was considered. Dark matter masses ranging from 200~GeV to 10~TeV and mediator lifetimes of 0.0001~s to 10~s were considered. The mediator mass is assumed to be much smaller than the dark matter particle mass. Sensitivities were computed for mediators decaying into muon neutrinos directly.\\ 
\indent The analysis method employed here is a likelihood based analysis. To distinguish signal from background, a region of interest around the position of the Sun is used. The region of interest is defined as events with an angular separation with respect to the Sun direction, $\theta$, such that $\theta < 11.5^\circ$. Then the likelihood function used in this analysis is defined as

\begin{equation}
    \mathcal{L}(n_s)= \prod_{i=1}^{N_{tot}}\left(n_s S(\psi_i,E_i) + N_{tot} B(\psi_i,E_i) \right) e^{-(n_s+N_{tot})},
\end{equation}

where $n_s$ is the supposed number of signal events, $N_{tot}$ is the total number of events in the sample and $S$ and $B$ are probability density functions (pdfs) describing  the likelihood of the event with an angular separation to the Sun $\psi_i$ and reconstructed energy $E_i$ for the event number $i$. To generate the signal pdf {\it S} the expected neutrino signal was taken from~\cite{bell}. The background pdf {\it B} is estimated using real data and scrambling the right ascension direction of the events. The likelihood is then optimized with respect to $n_s$ and a test statistic value (TS) is calculated as

\begin{equation}
    TS= -2\log\left(\frac{\mathcal{L}(n_{opt})}{\mathcal{L}(0)}\right),
\end{equation}

where $n_{opt}$ is the value of $n_s$ where the likelihood is maximal. To study the effectiveness of the likelihood function pseudo experiments (PEs) with varying amounts of inserted signal events were generated and from these TS distributions were calculated. 
Using the Feldman--Cousins method~\cite{fcousins} confidence intervals at a confidence level (CL) of 90\% are calculated for fixed simulated values of signal strength. Sensitivities are set at a signal strength for which the corresponding confidence interval lies entirely above the median of the TS distribution for PEs with no inserted signal.
The initial sensitivities are set on average numbers of detected events. These have to be converted to relevant parameters of secluded dark matter models. The conversion to neutrino flux sensitivities is done using the detector acceptance $Acc$, which is defined by:

\begin{equation}
    \Phi_{90\%}= \frac{\mu_{90\%}}{Acc},
\end{equation}

with the 90\% CL sensitivity to the total neutrino flux $\Phi_{90\%}$ and the 90\% CL sensitivity in term of average detected signal neutrino events $\mu_{90\%}$. The acceptance is calculated using the same neutrino spectra used to calculate the function $S$ and the IceCube detector simulation. In the next step conversion factors calculated from the DarkSUSY simulation package~\cite{conversion} are used to convert these into sensitivities to spin dependent and spin independent dark matter-nucleon scattering. 
In the conversion factor an equilibrium between annihilation and gravitational capture is assumed such that

\begin{equation}
    C_A= \frac{C_{C}}{2}
\end{equation}

for the annihilation rate $C_A$ and the capture rate $C_C$ holds. Such an equilibrium will establish itself in the Sun at a timescale 

\begin{equation}
    \tau= \frac{1}{\sqrt{C_C C_A}}.
\end{equation}

With an age of the Sun of $4.7 \times 10^9$ years this is generally the case for this type of search. Using the number of neutrinos per annihilation $N_\nu$ calculated from the spectra from~\cite{bell} for each mediator lifetime and dark matter mass the annihilation rate is calculated as

\begin{equation}
    C_A= \frac{4\pi \rm AU^2 \Phi_{90\%}}{N_\nu},
\end{equation}

where AU is the astronomical unit. Spin dependent scattering cross sections are then calculated using the results in the publication~\cite{conversion}.

\section{Sensitivities}\label{sec:Results}

\begin{figure}
	\centering
	\includegraphics[width=0.7\textwidth]{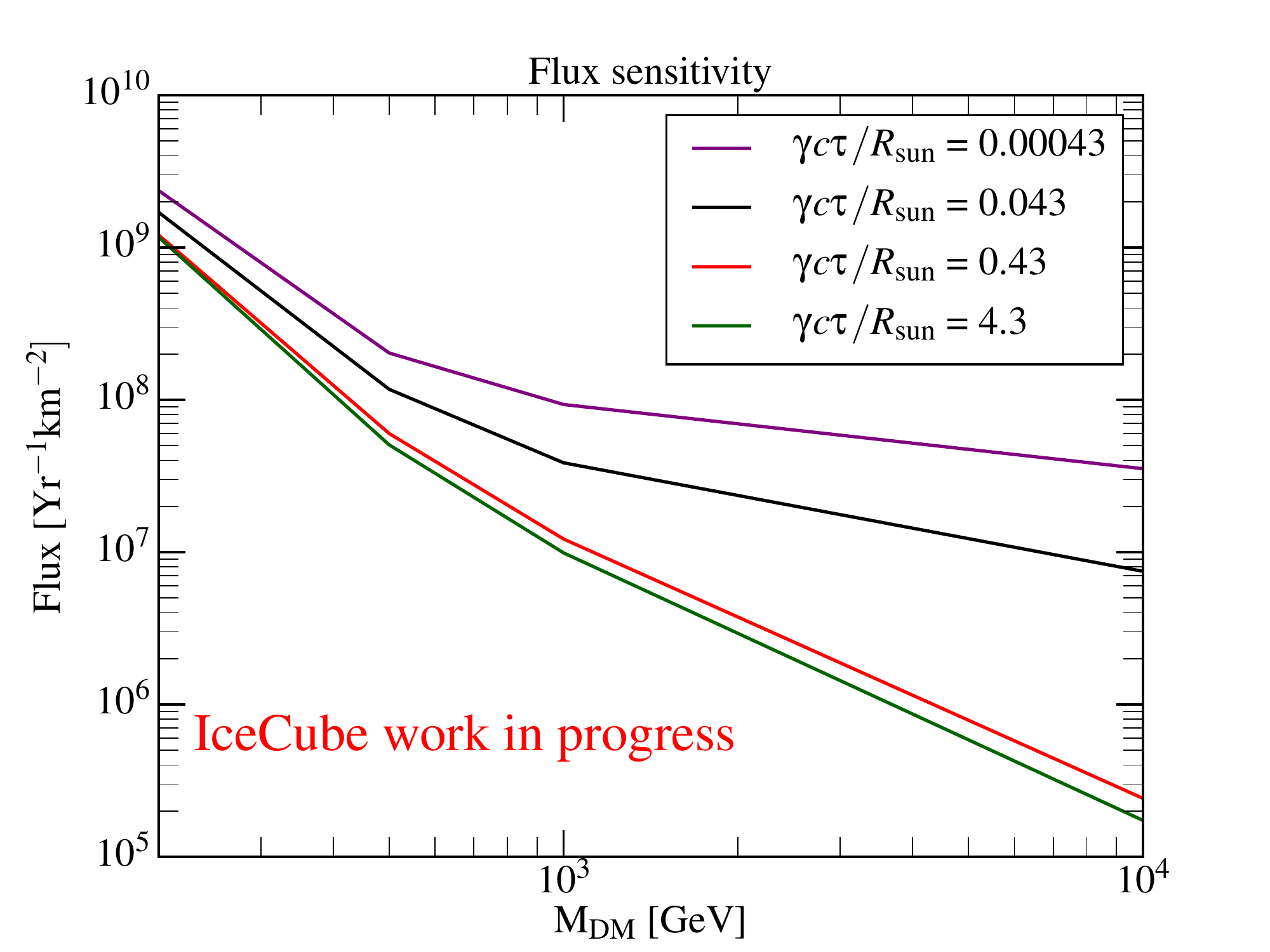}
	\caption{Sensitivities to the muon neutrino and antineutrino flux from secluded dark matter annihilations in the Sun with mediators decaying into muon neutrinos for different mediator lifetimes.}\label{fig:Flux}
\end{figure}

\begin{figure}
	\centering
	\includegraphics[width=0.7\textwidth]{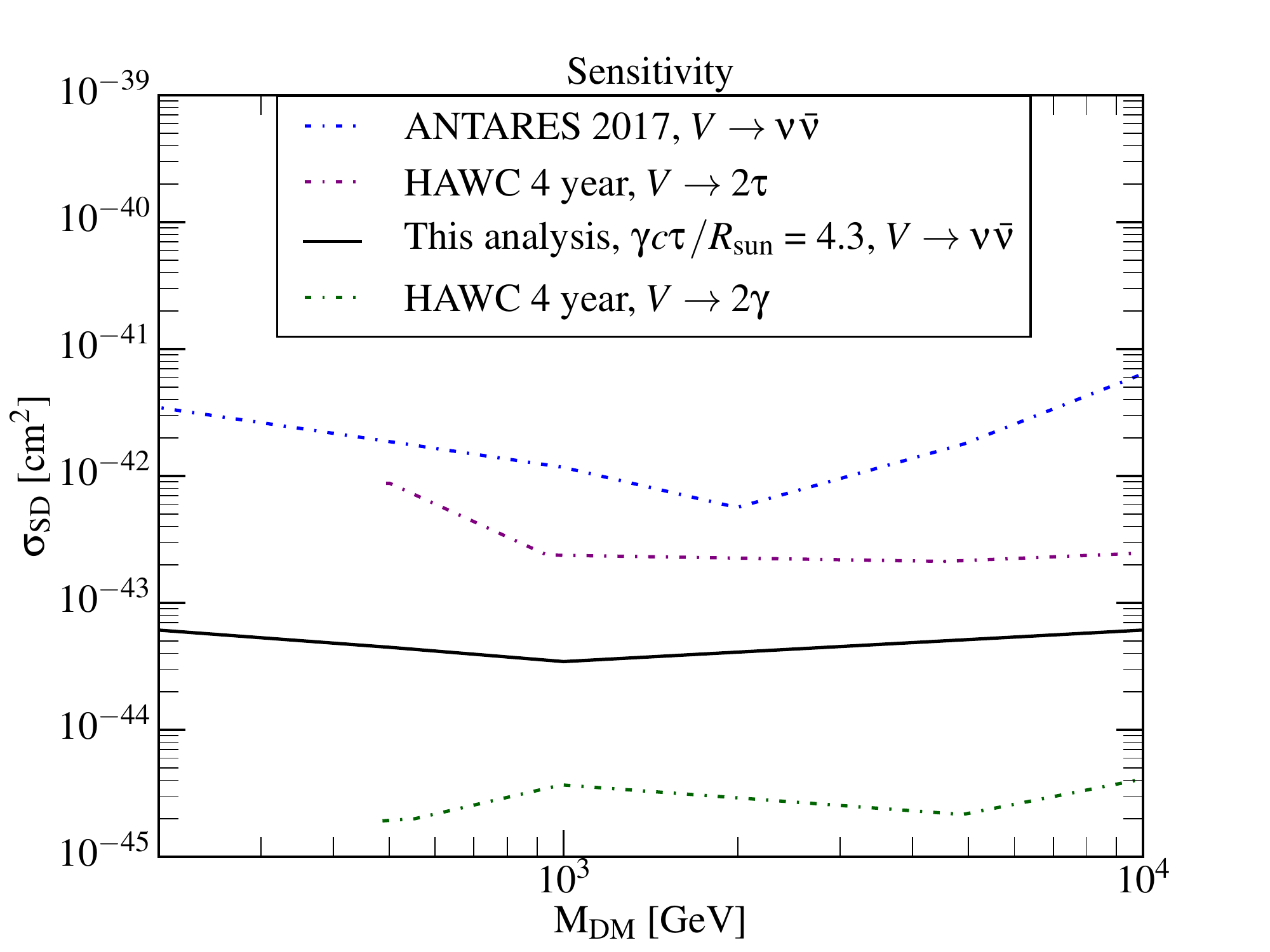}
	\caption{Sensitivities to the spin dependent dark matter scattering cross section in comparison to previous limits based on IceCube public data and limits from the HAWC experiment assuming the mediator decays into photons}\label{fig:SDCS}
\end{figure}

Sensitivities to the signal neutrino flux are shown in figure \ref{fig:Flux}. Here the dependency of the sensitivity on the mediator lifetime is visible. Sensitivities significantly improve at higher mediator lifetimes, especially at high dark matter particle masses. The final sensitivities to the spin dependent scattering cross section are shown in figure \ref{fig:SDCS}. The sensitivities are compared to previous results from an analysis using IceCube public data and to recent results from the HAWC experiment~\cite{HAWC}. In the previous analysis using IceCube data mediator lifetimes long enough to avoid interactions with the Sun were used. Consequently the sensitivities for a mediator lifetime of 10~s provide the most accurate comparison. It can seen that this analysis presents a significant improvement. This can be attributed to the use of a larger dataset, a more sophisticated analysis method and the use of the neutrino event energy. 

For the HAWC limits a mediator decaying purely into pairs of photons and tau leptons was assumed. The comparison to the limits for decays into photons of is disfavourable, while the limits for decays into tau leptons are surpassed. Neither channel allows a good comparison of the detectors themselves and given the very different decay channels IceCube results are complementary to other types of secluded dark matter searches such as those with gamma-ray experiments. A better comparison of the experiment could be done using mediators decaying into tau leptons for the IceCube detector. Such sensitivities are currently in preparation.

\section{Conclusion}

In conclusion this analysis shows a significant improvement over previous analyses and will be the first analysis of this kind performed by the IceCube collaboration. The expected sensitivities are highly complimentary to the results of other experiments and are the best sensitivities for this type of dark matter model from neutrino experiments for dark matter particles of masses between 200~GeV and 10~TeV. We show that IceCube is sensitive to spin--dependent dark matter--nucleon scattering cross--sections of $3.45 \cdot 10^{-34}~\rm cm^2$ for dark matter masses of 1 TeV. The analysis is close to be finished and the final results of this search are to be published in the future.

\bibliographystyle{ICRC}

\end{document}